\documentstyle[12pt,epsf,rotating,axodraw]{article}

\catcode`@=11
\def\citer{\@ifnextchar
[{\@tempswatrue\@citexr}{\@tempswafalse\@citexr[]}}

\def\@citexr[#1]#2{\if@filesw\immediate\write\@auxout{\string\citation{#2}}\fi
  \def\@citea{}\@cite{\@for\@citeb:=#2\do
    {\@citea\def\@citea{--\penalty\@m}\@ifundefined
       {b@\@citeb}{{\bf ?}\@warning
       {Citation `\@citeb' on page \thepage \space undefined}}%
\hbox{\csname b@\@citeb\endcsname}}}{#1}}
\catcode`@=12

\begin{document}

\renewcommand{\thefootnote}{\fnsymbol{footnote}}

\begin{flushright}
WUE-ITP-98-058 \\
hep-ph/9902296 \\
February 1999
\end{flushright}

\vspace*{1cm}

\begin{center}
{\Large \bf A Note on $W$ Boson Production at HERA\footnote{Contribution
to the 3rd UK Phenomenology Workshop on HERA Physics, Durham, 20-25
Sep. 1998.}} \\[0.5cm]
{\sc P.~Nason$^1$, R.~R\"uckl$^2$ and M. Spira$^3$} \\
$^1${\it INFN, Sezione di Milano, Milan, Italy} \\
$^2$ {\it Institut f\"ur Theoretische Physik, Universit\"at W\"urzburg,
D-97074 W\"urzburg, Germany} \\
$^3${\it II.\ Institut f\"ur Theoretische Physik, Universit\"at Hamburg,
D-22761 Hamburg, Germany} \\[0.5cm]
\end{center}
\begin{abstract}
We discuss $W$ boson production at HERA including NLO QCD corrections.
\end{abstract}

\vspace*{0.5cm}

The production of $W$ bosons at $ep$ colliders is mediated by photon, $Z$
and $W$ exchange between the electron/positron and the hadronic part of
the process \cite{baur}. In practice, it is convenient to distinguish two
regions, a small $Q^2$ one (photoproduction region), that can be treated
using the Weizs\"acker-Williams photon spectrum convoluted with the cross
section for $\gamma q \to q' W$, and a large $Q^2$ one (DIS region). 

While the treatment of the DIS region is straightforward [a typical
contribution is shown in the third diagram of Fig.~\ref{fg:1}], the small
$Q^2$ region requires the inclusion of the contribution of the hadronic
component of the photon, in which the photon, behaving like a hadron,
produces the $W$ in the collision with the proton via the standard
Drell-Yan mechanism [first diagram of Fig.~\ref{fg:1}]. This component is,
in fact, the dominant one for this process, and full NLO corrections
should include both the NLO correction to the Drell-Yan process
\cite{drellyan}, and the
leading hard photon contribution [a typical contribution is shown in the
second diagram of Fig.~\ref{fg:1}], suitably subtracted for collinear
singularities. 

In the following, we separate the DIS from the small $Q^2$ region using an
angular cut of 5$^o$ on the outgoing lepton, corresponding to a cut
$Q^2_{max}$ in terms of the initial lepton energy \cite{q2max}.

For the resolved part we have evaluated the QCD corrections in the
$\overline{\rm MS}$ scheme. Defining the cross section at the values
$\mu_R = \mu_F = M_W$ for the renormalization and factorization scales,
the QCD corrections enhance the resolved contribution by about 40\% for
$W^+$ and $W^-$ production. In order to demonstrate the theoretical
uncertainties, the renormalization/factorization scale dependence of the
individual contributions to the process $e^+ p \to W^+ + X \to \mu^+
\nu_{\mu} + X$ [including the branching ratio $BR(W^+\to \mu^+ \nu_{\mu})
= 10.84\%$] are presented in Fig.~\ref{fg:2} for HERA conditions.
It is clearly visible that the scale
dependence in the sum of direct and resolved contributions is
significantly reduced, once the NLO corrections to the resolved part are
included. The full curve presents the total sum of NLO resolved, LO direct
and the DIS LO contribution, i.e.\ the total $W^+$ production cross
section after including the QCD corrections. The residual scale dependence
is about 5\%. Since the remaining dependence on $Q^2_{max}$ is of the same
size, the total theoretical uncertainty is estimated to be less than about
10\%. \\

\noindent
{\bf Acknowledgements.} \\
We would like to thank G.\ Altarelli, U.\ Baur and D.\ Zeppenfeld for
helpful discussions.

\begin{figure}[hbt]
\begin{center}
\begin{picture}(120,80)(20,10)

\ArrowLine(0,80)(50,50)
\ArrowLine(50,50)(0,20)
\Photon(50,50)(100,50){3}{5}

\put(-15,78){$q$}
\put(-15,18){$\bar q$}
\put(105,48){$W$}

\end{picture}
\begin{picture}(120,80)(-10,10)

\ArrowLine(0,20)(50,20)
\ArrowLine(50,20)(50,80)
\ArrowLine(50,80)(100,80)
\Photon(50,20)(100,20){3}{5}
\Photon(0,80)(50,80){3}{5}

\put(-15,78){$\gamma$}
\put(-15,18){$q$}
\put(55,48){$q'$}
\put(105,78){$q'$}
\put(105,18){$W$}

\end{picture}
\begin{picture}(120,80)(-40,10)

\ArrowLine(0,80)(50,80)
\ArrowLine(50,80)(100,80)
\ArrowLine(0,20)(50,20)
\ArrowLine(50,20)(50,50)
\ArrowLine(50,50)(100,50)
\Photon(50,20)(100,20){3}{5}
\Photon(50,50)(50,80){3}{3}

\put(-15,78){$e$}
\put(-15,18){$q$}
\put(55,33){$q'$}
\put(55,63){$\gamma,Z$}
\put(105,78){$e$}
\put(105,48){$q'$}
\put(105,18){$W$}

\end{picture}  \\
\caption[]{\label{fg:1} \it Typical diagrams of $W$ boson production at
HERA: resolved, direct and DIS part.}
\end{center}
\end{figure}
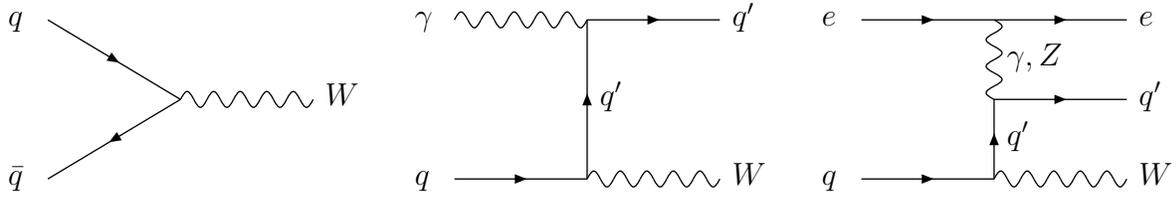
\begin{figure}[hbt]
\vspace*{0.5cm}

\hspace*{2.0cm}
\begin{turn}{-90}%
\epsfxsize=8cm \epsfbox{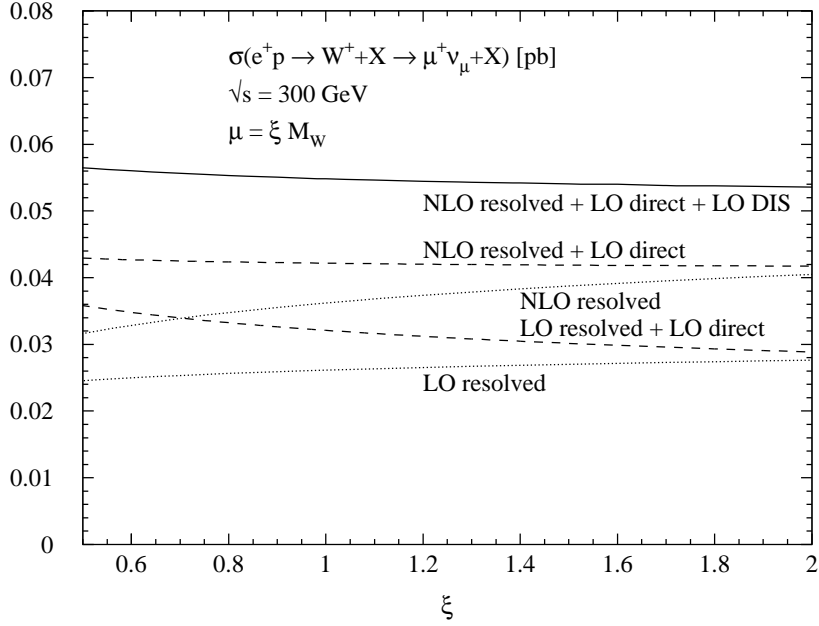}
\end{turn}
\vspace*{-0.0cm}

\caption[]{\label{fg:2} \it Renormalization and factorization scale
dependence [$\mu = \mu_F = \mu_R = \xi M_W$] for all individual
contributions. The full curve represents the final prediction for the
total cross section of $W^+$ production in $e^+p$ collisions. We have
chosen CTEQ4M and ACFGP parton densities for the proton and the photon,
respectively. The strong coupling constant is taken at NLO with
$\Lambda_5 = 202$ MeV. An angular cut of $5^o$ is introduced for the
separation of photoproduction and deep inelastic scattering (DIS).}
\end{figure}

\end{document}